\documentclass[twocolumn,nofootinbib,amsmath,amssymb]{revtex4}

\usepackage{graphicx}
\usepackage{dcolumn}
\usepackage{bm}
\usepackage{color}
\usepackage{cancel}
\usepackage{subfigure}

\def\be{\begin{equation}}
\def\ee{\end{equation}}
\def\bea{\begin{eqnarray}}
\def\eea{\end{eqnarray}}

\begin{document}

\title{What if Planck's Universe isn't flat?}

\author{Philip Bull}
%\email{phil.bull@astro.ox.ac.uk}
\affiliation{Department of Astrophysics, University of Oxford,
DWB, Keble Road, Oxford, OX1 3RH, UK} 

\author{Marc Kamionkowski}
%\email{kamion@jhu.edu}
\affiliation{Department of Physics and Astronomy, Johns Hopkins
University, Baltimore, MD\ 21218, USA}

\begin{abstract}
Inflationary theory predicts that the observable Universe should be 
very close to flat, with a spatial-curvature
parameter $|\Omega_K| \lesssim 10^{-4}$. The WMAP
satellite currently constrains $|\Omega_K| \lesssim 0.01$, and the
Planck satellite will be sensitive to values near $10^{-3}$. Suppose 
that Planck were to find $\Omega_K \!\neq 0$ at this level. 
Would this necessarily be a serious problem for inflation? We argue 
that an apparent departure from flatness could be due either to a 
local (wavelength comparable to the observable horizon) 
inhomogeneity, or a truly superhorizon departure from flatness.  If 
there is a local inhomogeneity, then secondary CMB anisotropies distort 
the CMB frequency spectrum at a level potentially detectable by a
next-generation experiment.  We discuss how these spectral
distortions would complement constraints on the Grishchuk-Zel'dovich
effect from the low-$\ell$ CMB power spectrum in discovering the 
source of the departure from flatness.
\end{abstract}

\maketitle

Inflation predicts that the observable Universe 
should be very nearly flat, with a spatial-curvature
parameter $|\Omega_K| < 10^{-4}$ in most models \cite{Linde2007}. 
WMAP data currently constrain $|\Omega_K| \lesssim 10^{-2}$ (95\% CL)
\cite{arXiv:1212.5226}, and Planck should be sensitive to
$\Omega_K$ at around the $10^{-3}$ level \cite{Knox:2005hx}, improving
to $\sim \!\! 10^{-4}$ when combined with 21-cm intensity maps
\cite{Mao:2008ug} (which represents the limit of detectability 
\cite{Vardanyan}).

Suppose that Planck were to find a nonzero value for $\Omega_K$.
What might this mean for inflation?  Such an observation would
nominally be evidence for a genuine departure from flatness on
superhorizon scales, with wide-ranging implications for a broad
class of inflationary models; for example, a measurement of
$\Omega_K < -10^{-4}$ is sufficient to rule out the majority of
eternal inflation scenarios with high confidence
\cite{Kleban:2012ph}.

Before jumping to such conclusions, though, one might wonder whether
the deviation could be explained simply by a local inhomogeneity 
that biases our determinations of cosmological parameters. This would 
allow us to preserve flatness (and thus some relatively natural sort of inflation) 
by explaining the discrepancy as the result of systematic distortions 
of, e.g., the distance-redshift relation due to lensing by the
inhomogeneity \cite{Valkenburg:2012ds}.  Although a local density 
fluctuation of a large enough amplitude ($\Phi \gtrsim10^{-3}$) would be 
inconsistent with the simplest inflationary models, it might conceivably 
arise if there is some strongly scale-dependent non-Gaussianity, or
perhaps if some sort of semi-classical fluctuation arises at the
beginning or end of inflation \cite{Aslanyan:2013zs}.

For a sufficiently large and smooth local inhomogeneity, it would be 
difficult to definitively distinguish these two situations using 
standard cosmological tests. Purely geometric observables such as 
distance measures would be inhibited by degeneracies with 
evolving-dark-energy models \cite{Valkenburg:2011ty}, and the 
deviation from flatness would be too small to significantly affect 
the growth of structure.

In this Letter, we show that a class of observables based on spectral 
distortions of the CMB offer the prospect to disentangle
the two scenarios. These observables exploit the strong relationship between 
spatial homogeneity and the isotropy of spacetime; by using them to 
measure the dipole anisotropy of the CMB about distant points, it is 
possible to place stringent constraints on the possible size of a local 
inhomogeneity \cite{Goodman:1995dt}. Furthermore, these 
observables unambiguously distinguish between subhorizon and 
superhorizon effects, owing to a cancellation of the dipole induced by 
superhorizon perturbations \cite{Grishchuk,Erickcek:2008jp}.

We begin by calculating the bias in $\Omega_K$ due to a local
inhomogeneity. We take the form of this local inhomogeneity
throughout to be a spherically symmetric potential perturbation
$\Phi(r,t)= D(t) \, a^{-1}(t)  \Phi_0 \exp[-(r/r_0)^2]$, where the
linear-theory growth factor is normalized to $D\!\!=\!\!1$ today.
The presence of a large, local inhomogeneity modifies the apparent 
distance to last scattering through a combination of lensing, 
integrated Sachs-Wolfe effect, gravitational redshift, and
Doppler shift.  Ref.~\cite{Bonvin:2005ps} derived a full expression 
for the (subhorizon) luminosity-distance perturbation $\delta d_L$ up 
to linear order in perturbations. (The observed 
distance $d_L(z) = \bar{d}_L (1 + \delta d_L)$, where the overbar 
denotes a background quantity.) When considering the CMB, it is useful 
to rewrite this as a perturbation $\delta d_A = \delta d_L - {2
\delta z}/(1+z)$ to the angular-diameter distance.

\begin{figure}[th]
\vspace{-0.35cm}
\hspace*{-0.5cm}
\includegraphics[width=1.15\columnwidth]{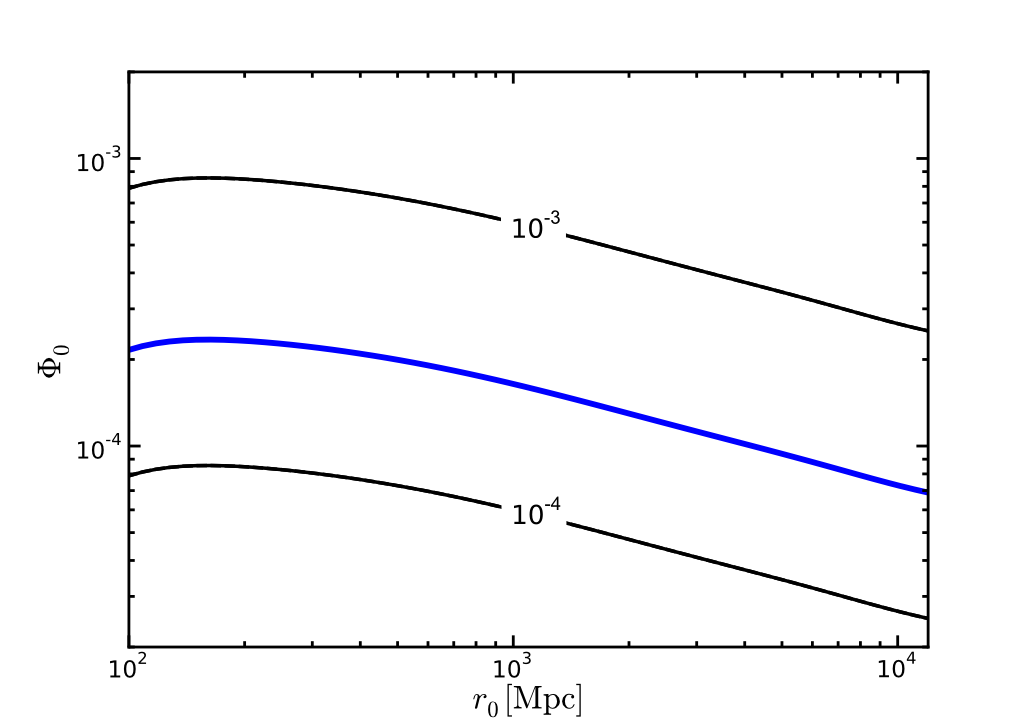}
\caption{The change $\delta d_A$ in the distance to last
 scattering as a 
 function of the width and depth of the inhomogeneity. Black
 lines denote $\delta d_A$.
 The thick blue line plots the difference in distance (in background)
 between models with $|\Omega_K|=10^{-3}$ and $\Omega_K=0$ (with 
 identical $h$ and $\Omega_m$), equivalent to 
 $\delta d_A = 2.7\times 10^{-4}$.}
\label{fig-shift}
\end{figure}

An observer sitting at the center of a spherically-symmetric 
inhomogeneity will measure a distance to last scattering which deviates 
from the background quantity by a uniform amount over the whole sky. 
This introduces a shift in angular scale of the 
entire CMB power spectrum. The value of $\Omega_K$ inferred from 
observations depends primarily on the angular scale of the first 
few CMB acoustic peaks \cite{Kamionkowski:1993aw}, and will
therefore be biased away from its 
background value. Fig.~\ref{fig-shift} shows the distance perturbation 
as a function of the depth and width of a local inhomogeneity, compared 
with the change in (background) distance between a flat model, and one 
with $|\Omega_K|=10^{-3}$.  (For numerical work we take $[h,
\Omega_m, \Omega_\Lambda,\sigma_8] = [0.71,0.266,0.734,0.8]$ and
redshifts of reionization and last scattering to be $z_\mathrm{re}=10$
and $z_*=1090.79$ respectively.)  Based on the distance to last scattering 
alone, an inhomogeneity with $\Phi_0 \sim 10^{-3}$ 
would induce an apparent shift in $\Omega_K$ of order $10^{-3}$ 
for a wide range of widths.

The inhomogeneity will also cause the observed redshift $z_*$ of
the surface of last scattering, to differ from its background value, 
$\bar{z}_* = z_* - \delta z_*$, where \cite{Bonvin:2005ps}
% \bea \label{eqn-perturbed-z}
%      \delta z = (1 + z_S) \bigg [ \Phi_S - \Phi_O +
%      (\mathbf{v}_O - \mathbf{v}_S) \cdot \mathbf{n}~
%      \nonumber\\
%      \left . + 2 \int_{\eta_S}^{\eta_O} d\eta~ \mathbf{n} \cdot
%      \nabla \Phi\right ],
% \eea
\be \label{eqn-perturbed-z}
     \delta z = (1 + z_s) \left [ \Phi_s - \Phi_o +
     (\mathbf{v}_o - \mathbf{v}_s) \cdot \mathbf{n}
        + 2 \int_{\eta_s}^{\eta_o} d\eta~ \mathbf{n} \cdot
     \nabla \Phi\right ],
\ee
and $\mathbf{n}$ is a unit vector along the line of sight.
For the central observer, the effect of the redshift perturbation is to 
change the inferred conformal time (and thus expansion rate) of last 
scattering, which will bias the estimation of parameters such as 
$\Omega_m$. For an observer who is {\it off-center} in the 
inhomogeneity, however, an additional anisotropy will also be induced 
in the CMB. This is because the redshift perturbation,
Eq.~(\ref{eqn-perturbed-z}), depends on direction; a line of
sight looking 
towards the center of the inhomogeneity will experience a different 
change in redshift to one looking away from it, and thus there will be 
a direction-dependent change in temperature.

In general, anisotropies will be induced over a range of angular 
scales, but at least for observers close to the center of a large (wide) 
inhomogeneity, the dipole, $\beta$, will dominate. While there is also 
a dipole contribution due to the peculiar velocity of the observer, 
velocity perturbations due to the matter distribution on smaller scales 
are expected to be Gaussian random distributed with mean zero, whereas 
the dipole due to a large inhomogeneity will generally present a 
{\it systematic} trend in redshift and angle on the sky. This allows 
us to distinguish between the two contributions.
For a spherical inhomogeneity, axial symmetry dictates that the dipole 
will be aligned in the radial direction, and that all
spherical-harmonic modes of the induced anisotropy with $m \neq
0$ on the sky of the observer will be zero, so that $\beta
\propto \int \delta z_*(\theta) \cos\theta\sin\theta d\theta$.

We now discuss spectral distortions due to a local inhomogeneity.
There is a close relationship between homogeneity and the isotropy of 
spacetime. A number of observational tests that are sensitive to CMB 
anisotropies about distant points can be used to exploit this link and 
detect local inhomogeneities of the kind that would cause a systematic 
bias in measurements of $\Omega_K$.

The strength of the connection between homogeneity and isotropy is most clearly 
demonstrated by the Ehlers-Geren-Sachs (EGS) theorem \cite{Ehlers:1966ad}. 
According to EGS, if all comoving observers 
in a patch of spacetime see an isotropic CMB radiation field, then that 
patch is uniquely FLRW (i.e., it is necessarily homogeneous and isotropic). 
Generalizations of this result show that it is perturbatively stable, in the 
sense that small departures from perfect isotropy imply only small departures 
from homogeneity (up to some assumptions; see \cite{Rasanen} for a critique). 
A corollary to the EGS theorem is that observers in an inhomogeneous region 
of spacetime will in general see an anisotropic CMB sky. We can therefore use 
measurements of the anisotropy of the CMB about a collection of spacetime 
points to constrain the degree of inhomogeneity inside our Hubble volume 
\cite{Clifton:2011sn}.

Compton scattering of CMB radiation by ionized gas provides a way 
to detect anisotropy about remote points. The scattered radiation 
spectrum consists of a weighted superposition of spectra from all 
directions on the scatterer's sky, 
$I^\prime_\nu \sim \int \tau (1 + \cos^2\theta) I_\nu(\theta, \phi) d\Omega$.
If the scatterer's sky is a perfectly isotropic blackbody of 
uniform temperature, the scattered spectrum is simply a blackbody of 
the same temperature, plus spectral distortions due to the random 
thermal motions of the electrons in the scattering medium (the thermal 
Sunyaev-Zel'dovich effect, TSZ \cite{Sunyaev:1970eu}). If its sky is anisotropic, 
however, the resulting spectrum is a combination of blackbodies of different 
temperatures. This induces additional blackbody spectral distortions, 
and shifts the temperature of the `base' blackbody spectrum as seen by 
an observer \cite{Goodman:1995dt,Chluba:2004cn}. If
the dipole anisotropy dominates, we call these
the Compton-$y$ distortion and the kinematic Sunyaev-Zel'dovich
(KSZ) effect \cite{Sunyaev:1980nv}, respectively.

By measuring the Compton-$y$ distortion and KSZ effects for many 
scattering regions on our own sky, we can build up a picture of the 
degree of anisotropy, and thus inhomogeneity, within our past lightcone. 
We will now outline three observational tests based on these effects, and 
estimate their sensitivities to a local inhomogeneity.

We begin with the KSZ effect from galaxy clusters.
Galaxy clusters contain a significant amount of ionized gas. Since they 
are effectively individual collapsed objects, they can be used to 
sample the dipole anisotropy induced by a local inhomogeneity 
at discrete points in space. This is useful to reconstruct the 
systematic trend in dipole anisotropy as a function of redshift that 
a local inhomogeneity produces. Each cluster has a 
characteristic integrated optical depth of $\tau \sim 10^{-3}-10^{-2}$. 
The KSZ signal due to a single galaxy cluster at redshift $z$ is 
$\Delta T/T = - \beta(z) \tau$, and can be extracted from CMB sky 
maps given a sufficiently accurate component separation method and 
low-noise data.

The KSZ effect from individual clusters is difficult to measure owing 
to the smallness of the signal, confusion with primary CMB anisotropies, 
and other dominant systematic errors. Currently, only upper limits are 
available, but this is likely to change as data from Planck and 
small-scale CMB experiments such as ACT and SPT become available. 
Current data have nevertheless been used to constrain inhomogeneous 
relativistic cosmological models for dark energy \cite{GarciaBellido:2008gd}.

\begin{figure}[t]
%\hspace{-1.1cm}
\vspace{-0.3cm}
\hspace*{-0.6cm}
\includegraphics[width=1.15\columnwidth]{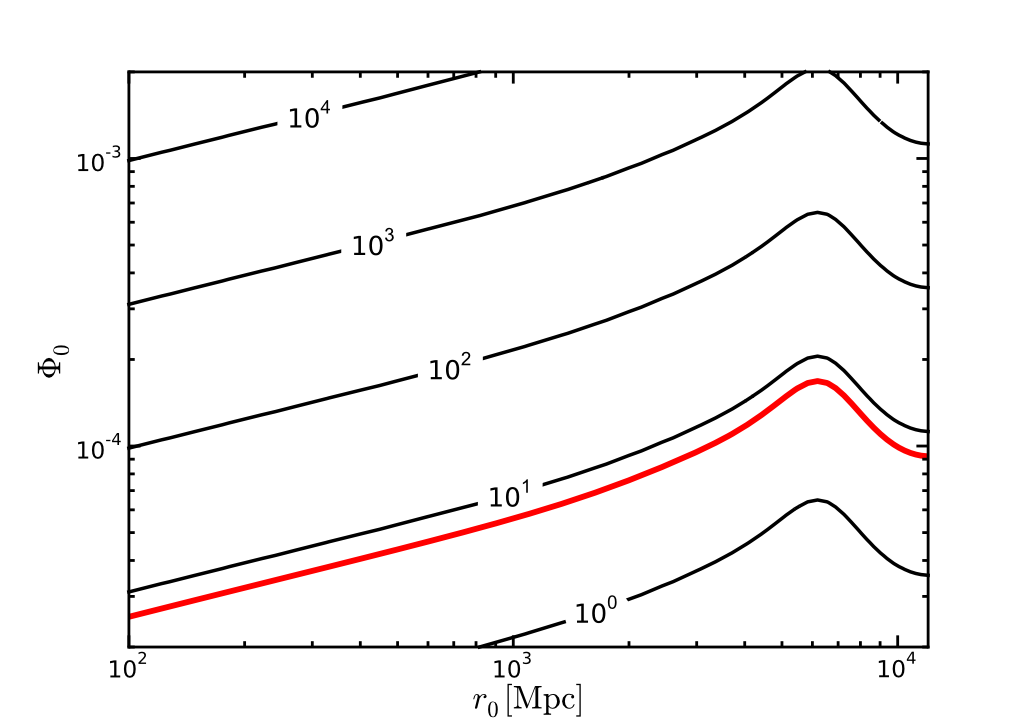}
\caption{The KSZ power $D_\ell = \ell(\ell+1)C_\ell T_0^2/2\pi$ at 
$\ell=3000$, in $\mu$K$^2$. The thick red line is the SPT upper limit 
of $D_{3000} < 6.7 \mu$K$^2$ (95\% CL).}
\label{fig-ksz-power}
\end{figure}

We now consider the KSZ angular power spectrum from gas in the
intergalactic medium.  This angular power spectrum is easier to
measure than the KSZ effect from individual clusters
because it is an integrated quantity and has 
additional contributions from the diffuse intergalactic medium that is 
not associated with clusters (sometimes called the Ostriker-Vishniac 
effect \cite{OstrikerVishniac}). The KSZ power spectrum from a
large inhomogeneity is \cite{Moss:2011ze}
\be
     C_\ell \approx {8 \pi^3} \int_0^{r_\mathrm{re}} dr ~r^{-3}
     \left [ \beta(z) (d\tau/dr) \right ]^2 P(k(r),z(r)).
\ee
The Limber approximation has been used, giving 
$k(r) = (2\ell+1)/2 r(z)$.  We model the distribution of
scatterers in the late Universe with $d\tau/dz \propto \sigma_T
f_b \rho(z)/H(z)$ \cite{Moss:2011ze}, and take reionization
to be an abrupt transition at $z_{\mathrm{re}}$. 
At high $\ell$, the KSZ signal is strongly dependent 
on the non-linear matter power spectrum, $P(k, z)$, which we model 
using HaloFit/CLASS \cite{Blas:2011rf}.  Results for our toy model 
are shown in Fig.~\ref{fig-ksz-power}.

At a characteristic angular scale of $\ell \sim 3000$ (where the 
primary CMB signal becomes subdominant), the signal is dominated by 
contributions from small-scale matter inhomogeneities at lower redshifts, where 
the induced anisotropy is mostly dipolar. These scales are accessible 
to CMB experiments such as ACT and SPT, which have recently put stringent 
upper limits on the combined TSZ+KSZ power \cite{Sievers:2013wk}.
Accessing the bare KSZ signal is complicated by difficulties in 
modeling the distribution of extragalactic point sources \cite{Addison:2012my}, 
and contains a theoretical uncertainty due to the unknown `patchiness' of 
reionization, which also contributes a KSZ effect \cite{Santos:2003jb}.

\begin{figure}[t]
\vspace{-0.25cm}
\hspace*{-0.5cm}
\includegraphics[width=1.15\columnwidth]{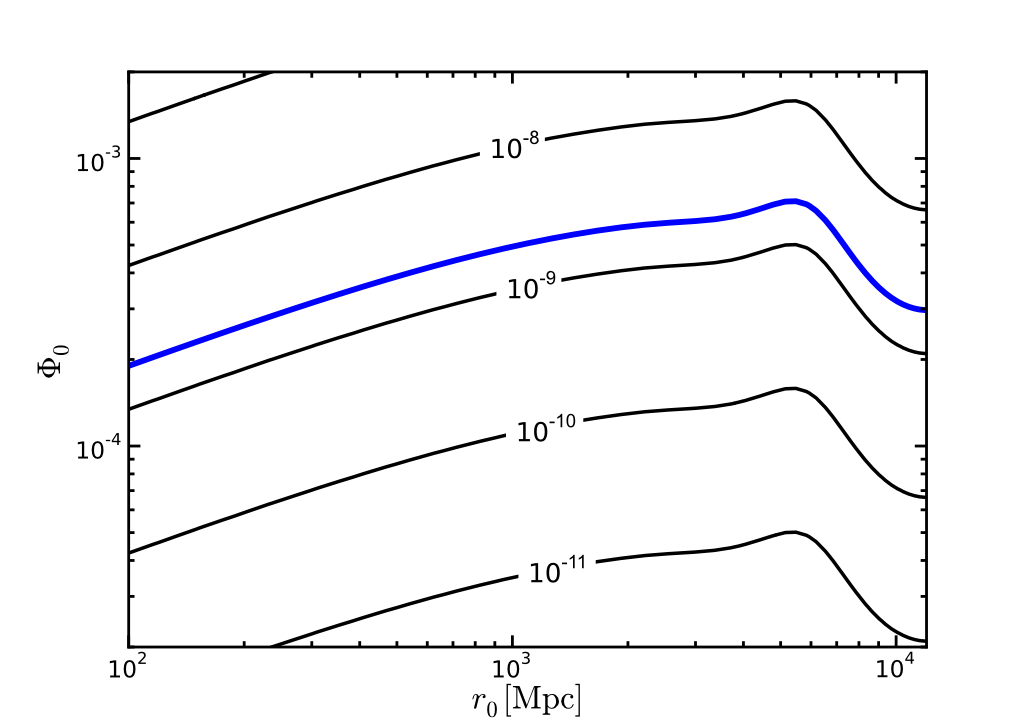}
\caption{The Compton-$y$ distortion induced by the 
 inhomogeneity. Also plotted is the projected upper limit from 
 PIXIE (thick blue line).}
\label{fig-y-distortion}
\vspace{-0.5cm}
\end{figure}

We now turn to the Compton-$y$ distortion induced by the
inhomogeneity.  Spectral distortions arising from the Compton
scattering of an anisotropic CMB can be parametrized as a
Compton-$y$ blackbody distortion.  When the dipole
dominates, the observed Compton-$y$ distortion is a monopole
\cite{Moss:2011ze},
\be
     y = (7/10) \int_0^{r_\mathrm{re}} dr ~
     (d\tau/dr) \beta^2(r).
\ee
Results for our model are shown in Fig. \ref{fig-y-distortion}.

Measurement of the Compton-$y$ distortion requires an instrument for which an 
absolute calibration of the spectral response can be obtained. This 
excludes most recent CMB experiments, and so the best current 
constraints come from COBE/FIRAS \cite{COBE}. The planned PIXIE 
mission \cite{Kogut:2011xw} could improve the determination of $y$ by some 
four orders of magnitude.

Our toy-model calculations give some sense of the effectiveness of the 
different spectral-distortion tests in constraining the size of a local 
inhomogeneity. A depth of $\Phi_0 \sim 2 \times 10^{-4}$ is sufficient 
to induce a bias in the inferred spatial curvature of 
$\Delta \Omega_K \approx 10^{-3}$ for a wide range of $r_0$ 
(Fig. \ref{fig-shift}). Existing upper limits on the KSZ power at 
$\ell=3000$ from SPT are sufficient to rule out an inhomogeneity of 
this depth with a width less than around 5 Gpc, although larger $r_0$ 
are still allowed (Fig. \ref{fig-ksz-power}). The Compton-$y$
distortion, on the other hand, provides much weaker constraints 
even with the great increase in precision that would be possible with 
PIXIE (Fig. \ref{fig-y-distortion}).  Part of the reason for the 
relative effectiveness of the KSZ power spectrum is its density 
weighting, which enhances the signal at the high $\ell$ probed by 
precision CMB experiments.

The above calculations are only intended to be illustrative, and more 
detailed modeling would be required to produce firmer constraints. For 
example, the KSZ angular power spectrum is sensitive to the non-linear 
contributions to $P(k)$ \cite{Moss:2011ze}, and the form of $\beta(z)$ 
to the shape of the potential, $\Phi(r)$, so uncertainties in these 
functions should be treated carefully. For wider inhomogeneities, 
there is also a (relatively minor) dependence on the details of 
reionization. Finally, the assumption that the inhomogeneity is perfectly 
spherically symmetric, and that we are exactly at its center, should 
also be relaxed.  A realistic inhomogeneity cannot be too 
asymmetric, or place us too far from the center, however,
without violating limits on isotropy, moderating a CMB dipole
that is observed locally \cite{Foreman:2010uj}, or inducing a CMB
statistical anisotropy \cite{Aslanyan:2013zs}.

% Where would such an inhomogeneity come from in the first place?
Why should we expect to find ourselves near to the center of a large 
inhomogeneity in the first place? Although such a situation may seem 
unlikely \cite{Foreman:2010uj}, there are inflationary mechanisms known in 
the literature which preferentially place observers near the center of 
large underdensities \cite{Linde:1996hg}. Furthermore, Ellis \cite{Ellis} 
has argued that it would be inconsistent to rule out such inhomogeneities 
on strictly {\it a priori} probabilistic grounds, since we currently 
accept features in our cosmological models that are substantially less 
probable anyway. As such, observations should be the final arbiter in 
deciding whether a large inhomogeneity exists or not.

% Would other, more obvious, effects arise?
Wouldn't its presence have already been discovered through other 
observational probes? Inhomogeneities of the kind considered here 
modify the low-$\ell$ CMB, causing alignment of low-$\ell$ multipoles 
\cite{Alnes:2006pf}, and changes in the ISW signal, temperature-polarization 
cross-spectrum, and associated modifications to the reionization 
history \cite{Moss:2010jx}. Unfortunately, the induced effects tend either to 
be smaller than cosmic variance at the relevant scales, or strongly 
dependent on the details of the model, rendering these tests 
inconclusive.

% Discussion of the Grishchuk-Zel'dovich effect
Superhorizon perturbations also produce fluctuations in the low-$\ell$ 
CMB through the Grishchuk-Zel'dovich effect \cite{Grishchuk}. In a 
number of cosmological models (including $\Lambda$CDM), it has been 
shown that there is a cancellation between the anisotropy and peculiar 
velocity induced by such perturbations, resulting in no net dipole to 
first order \cite{Erickcek:2008jp}. Constraints from the low-$\ell$ CMB are 
therefore complementary to spectral-distortion tests of the sort 
outlined above: A deviation from spatial flatness caused by a local 
inhomogeneity results in a net dipole about many locations within our 
horizon, which can be measured using, e.g., the KSZ effect, whereas a 
superhorizon deviation from flatness will produce no such signal, 
instead causing an enhancement of the quadrupole and higher moments of 
our local CMB.

% Conclusion
In conclusion, an observation of $|\Omega_K| \gtrsim 10^{-4}$
would have considerable implications for inflation but would
not, on its own, be sufficient to rule out eternal inflation.  It would
also have to be shown that the inferred deviation from flatness
was not caused by the effects of a local inhomogeneity
instead. Observations of CMB spectral distortions such as the
KSZ effect and Compton-$y$ distortion, taken with constraints
on the size of the Grishchuk-Zel'dovich effect from the
low-$\ell$ CMB power spectrum, present a viable method to
constrain the source of a seeming departure from flatness.

We thank P.\ G.\ Ferreira for useful discussions. PB
acknowledges the support of the STFC.  MK was supported by DoE
SC-0008108 and NASA NNX12AE86G. The computer code used in this paper 
is available online at \url{www.physics.ox.ac.uk/users/bullp}, and 
makes use of the CosmoloPy package (\url{http://roban.github.com/CosmoloPy}).

{\bf Note added:} An error affecting the figures in the original article was 
discovered after publication. An erratum has been published in Phys. Rev. D, 
and the corresponding corrections have been applied here. We are 
grateful to R. Durrer for pointing out an issue which led to the discovery of 
this error.

% ---------------------- BIBLIOGRAPHY -----------------------------------------
%\section*{References}

\end{document}